\documentclass[showpacs,prb,preprint]{revtex4}
\usepackage{graphicx}

\input{tcilatex}

\begin{document}

\title{Electrically Controlled Magnetic Memory and Programmable Logic
based on Graphene/Ferromagnet Hybrid Structures}
\author{Y. G. Semenov, J. M. Zavada, and K. W. Kim}

\begin{abstract}
It has been shown that the combining of the electrical effect on the
exchange bias field with giant magneto-resistance effect of the
graphene/ferromagnet hybrid structures reveals a new non-volatile
magnetic random access memory device conception. In such device an
electric bias realizes the writing bits instead a magnetic field of
remote word line with high energy consumption. Interplay of two
graphene mediated exchange bias fields applied to different sides of
free ferromagnet results in programable logic operations that
depends on specific realization of the structure.
\end{abstract}

\pacs{73.21.-b,85.75.-d,73.43.Qt,73.61.Wp}

\address{Department of Electrical and Computer Engineering, North
Carolina State University, Raleigh, NC 27695-7911}

\maketitle

\bigskip

Up-today a magnetic field driving\ by the current of remote write line
realizes the reversal of free ferromagnetic layer in a magnetic random
access memory (MRAM). The giant magnetoresistance (GMR) effect is used to
perform the reading in the storage stripe that is separated from word lines
writing the bits\cite{Tang95} or integrated with reading channel.\cite%
{Melo97} Along with MRAM the programmable logic can be designed on basis of
the GMR effect that utilizes high enough magnetic fields created by the
current through a programming line.\cite{Black00} A programmable computing
can be realized in such structures as a result of interplay between magnetic
fields of input lines affected top and bottom ferromagnetic FM layers with
different coercive fields.\cite{Ney03}

Some advantage in device scaling has another mechanism of the MRAM switching
based on spin transfer (or spin-torque)\cite{Slonczewski96,Berger96} in
pillar structures.\cite{Katine00,Jiang04} It was demonstrated that
spin-polarized current through the magnetic tunnel junction can drive the
magnetic switching in nanoscale device.\cite{Fuchs04} Note that both
aforementioned mechanisms of magnetization reversal need in high critical
switching current and consequently high energy dissipation accompanied
device operation. Design and creation of a composite free FM layer can
result in reducing spin transfer current\cite{Meng06} but it is not so
radically that waives the problem of large energy consumption.

At the same time several mechanisms, which completely do not relate to
electric current for magnetization reversal have been discussed in
literature. Magnetization spontaneous reversal when a temperature variation
modifies the exchange bias field in the magnetic heterostructures under the
external magnetic field was reported in Ref.\cite{Li06}. Involving a
multiferroic film to the magnetic structure brings about\ electrical control
of exchange bias field, as it has been recently discussed in Ref.\cite{Bea08}%
. Very recently the authors showed that an atomic thin graphite (graphene)
placed between two ferromagnetic dielectric layers (FDLs) realizes an
indirect exchange interaction between them and this interaction can be
easily controlled by applied electrical bias.\cite{SZKcondmat08} As this is
a case, it raises a natural question: Can an graphene-incorporated structure
accomplish the magnetization reversal of free FDL and how this effect can be
utilized in MRAM and logic devices?

In present study we analysis a different approach to the problem of
low-power-consuming non-volatile MRAM and logic design that based on the
unique properties of the graphene placed in interface between two magnetic
dielectric layers. In particular, the structure under consideration consists
of a three ferromagnetic dielectric layers (FDLs) constructed from same
material, which are coupled through monolayer graphene (MG) and bilayer
graphene (BG) (Fig. 1). The magnetization of the bottom ($\mathbf{M}_{b}$)
and top ($\mathbf{M}_{t}$) FDLs is pinned by strong enough coercivity or
proximate antiferromagnets in the usual fashion.\cite{Nogues99} By
technology reason they are assumed to be possessed the common direction
along $x$ axis so that $\mathbf{M}_{b}=\mathbf{M}_{t}\equiv \mathbf{M}_{0}$.
The magnetization $\mathbf{M}_{f}$ of the middle FDL can be controlled by
exchange bias fields mediated by monolayer ($\mathbf{H}_{1}$) and bilayer ($%
\mathbf{H}_{2}$) graphenes if their sum exceeds the coercivity of free FDL.

The origin of the $\mathbf{H}_{1}$ and $\mathbf{H}_{2}$ stems from the
exchange interactions $\alpha \mathbf{M}_{0}\mathbf{S}$ and $\alpha \mathbf{M%
}_{f}\mathbf{S}$ of graphene electrons with both pinned $\mathbf{M}_{0}$ and
free $\mathbf{M}_{f}$ FDLs (coefficient $\alpha $ is proportional to
electron-magnetic ion exchange integral, $\mathbf{S}$ is an electron spin).
This establishes an indirect interaction through the graphene electrons in
form of Zeeman energy $-\mathbf{H}_{n}\mathbf{M}_{f}$ ($n=1,2$).\cite%
{SZKcondmat08}

In very general way the thermodynamic potential $\Omega _{n}$ of the
graphene electrons interacted with both proximate FDLs determines the
effective magnetic field $\mathbf{H}_{n}=-\partial \Omega _{n}/\partial
\mathbf{M}_{f}$. Straightforward calculations show that $\mathbf{H}%
_{n}\parallel \mathbf{M}_{0}$ while the effective field projection on this
direction $H_{n}=\mathbf{H}_{n}\mathbf{M}_{0}\mathbf{/}\left\vert \mathbf{M}%
_{0}\right\vert $ is proportional to the modulation crest of the $\Omega
_{n} $ calculated for the electrons exchange interacted with proximate FDLs
with parallel, $\Omega _{n}^{(+)}=\Omega _{n}|_{\mathbf{M}_{f}\mathbf{=M}%
_{0}}$, and antiparallel, $\Omega _{n}^{(-)}=\Omega _{n}|_{\mathbf{M}_{f}%
\mathbf{=-M}_{0}}$ magnetization. Finally it can be shown\cite{SZKcondmat08}
that
\begin{equation}
H_{n}=\frac{\Delta \Omega _{n}}{2\mathcal{M}_{F}}\,,  \label{f9}
\end{equation}%
\ where $\Delta \Omega _{n}=\Omega _{n}^{(-)}-\Omega _{n}^{(+)}$, $\mathcal{M%
}_{f}=M_{f}A_{f}t_{f}$ is a total magnetic moment of the free FDL, $A_{f}$
and $t_{f}$ are its area and thickness. As in the case of conventional
exchange bias, the strength of $H_{n}$ is inversely proportional to the
thickness $t_{F}$ \cite{Nogues99}. This means that the influence of a back
magnetic field generated by free DFL on the top and bottom magnetic layers
can be strictly weaker their coercive fields under appropriate ferromagnetic
layers widths.

The dependence of $H_{n}$ [Eq. (\ref{f9})] on the electronic properties of
the graphene layer leads to qualitatively different characteristics for MG
and BG that the calculation of $\Delta \Omega _{1}$ and $\Delta \Omega _{2}$
highlights. In particular the carrier concentration (or position of the
electro-chemical potential $\mu $) variation by impurity doping or/and the
gate bias ($V_{g1}$, $V_{g2}$; see Fig.~1) significantly influences exchange
bias field in different manner for MG and BG.

Firstly, the signs of $H_{1}$ and $H_{2}$ are different in the wide range $%
\mu $. While MG tends to establish $\mathbf{M}_{f}$ parallel to $\mathbf{M}%
_{0}$ ($H_{1}>0$), BG favors the antiparallel alignment ($H_{2}<0$).
Secondly, a shift of $\mu $ from the graphene i-type point ($\mu =0$)
affects the strengths of the exchange bias fields $\mathbf{H}_{1}$ and $%
\mathbf{H}_{2}$ in the opposite directions. Namely, the magnitude of $H_{1}$
gradually increases with $|\mu |$ or electron/hole concentration, whereas
that of $\left\vert H_{2}\right\vert $ is at the maximum at $\mu =0$ with
minimal free carrier concentration and decreases to zero when $\left\vert
\mu \right\vert $ is large enough. The aforementioned characteristics can be
captured by an expression in terms of dimensionless electron exchange energy
$G=\alpha M_{0}/\gamma _{1}$ ($\gamma _{1}=0.4$ eV) and the factor $%
f_{n}(\mu )$, which provides the specific dependence on $\mu $ for MLG ($n=1$%
) and BLG ($n=2$),
\begin{equation}
\Delta \Omega _{n}=\frac{A_{F}}{a_{g}}G^{2}f_{n}(\mu ),  \label{f10}
\end{equation}%
where $a_{g}=0.0537$ nm$^{2}$ is the area of graphene primitive cells.
Hereinafter the temperature assumes to be 300 K.

Figure~2 shows the $f_{1}(\mu )$, $f_{2}(\mu )$ and their sum
vs.$\left\vert \mu \right\vert $\ evaluated at room temperature. As
evident from the figure, the shift of $\mu \simeq \pm 0.15\gamma
_{1}$ can change the strength of $H_{n}$ by about a factor of two
for both cases mediated by MG and BG. Moreover, the total field
strength $H_{1}+H_{2}\propto f_{1}(\mu
)+$ $f_{2}(\mu )$ that controls the magnetization of free FDL varies with $%
\mu $ from negative maximal magnitude through zero to positive value of
similar strength provided the $\mu $ equality for both MG and BG. The
strength of $H_{n}$ hase been estimated at actual $\mu $ and $G=0.1$ as $%
h/t_{f}$, where $h\approx 1000$ Oe$\cdot $nm.\cite{SZKcondmat08} Such
behavior inspires to use the joint action of $H_{1}$ and $H_{2}$ for
electrical switching of free FDL between two stable states, which correspond
to $\mathbf{M}_{f}$ parallel or antiparallel to $\mathbf{M}_{0}$. By analogy
with a magnetoresistive memory based on the "spin-valve" effect,\cite%
{Melo97,Matsuyama97} we explore the coercivity $H_{c}$ that secures the
stability of the magnetization with respect to quantum and thermal
fluctuations. Apparently the strength of coercive magnetic field must be
limited by inequality $H_{c}<h/t_{f}$. Indeed, in such a case there is a
range of exchange bias fields which, guarantees the reversal of
magnetization $\mathbf{M}_{f}$ while $\mathbf{M}_{0}$ remains with former
orientation.

The Fig. 2 also shows that the neutral charge point can correspond to the
effective fields compensation, $H_{1}+H_{2}=0$, if the equal impurity doping
of MG and BG with magnitude $\mu _{0}=0.11\gamma _{1}$ has been achieved.
Starting with this point, the total field can be variable in both directions
by depletion of free carriers or their extra population with
electro-chemical potential variation by applied electrical bias of both
polarity. Hereinafter the properly impurity doping is assumed to be done for
both graphene layers.

Once the problem of $\mathbf{M}_{f}$ switching between two states $\mathbf{M}%
_{f}=\mathbf{M}_{0}$ and $\mathbf{M}_{f}=-\mathbf{M}_{0}$ has been solved,
it can be utilized in non-volatile memory provided different magnetic states
of $\mathbf{M}_{f}$ are surely discerned with graphene electrical
properties. As it was recently shown, the conductivity of BG is
characterized by significant sensitivity with respect to magnetic ordering
of proximate FDLs because an misalignment of $\mathbf{M}_{f}$ and $\mathbf{M}%
_{0}$ results in bandgap opening and dispersion law
flattering.\cite{SZK08} Other possibility consists in magnetic state
detection through the MG electronic properties. As in the case of
BG, one can expect the higher conductivity in parallel configuration
than that in antiparallel
orientation. This is because band spin splitting at $\mathbf{M}_{f}=\mathbf{M%
}_{0}$ makes a finite electronic density of states at any electron energy
including Dirac point while this is not a case when graphene electrons do
not experience spin splitting at $\mathbf{M}_{f}=-\mathbf{M}_{0}$. We
calculate magnetoresistance of MG and compare it with that for BG at Fig.3.
The figure indicates some advantage of BG with respect to magnetoresistance
of MG that hase been depicted in device design (Fig. 1) with a BG reading
line.

At large enough negative $\Delta \mu \equiv \mu -\mu _{0}=\Delta \mu _{1}$
(Fig.4a) graphene-mediated field $\left\vert H_{1}+H_{2}\right\vert $
exceeds the $H_{c}$ and turns $\mathbf{M}_{f}$ opposite to $\mathbf{M}_{0}$
direction (Fig. 4b), i.e. their magnetization is antiparallel. Such
configuration corresponds to large graphene resistance $R_{1}$ (Fig.4c).\cite%
{SZK08} Such a state remains since the electrical biases will switch off
(Fig.4d). As soon as the chemical potentials of both graphene layers
supplies the positive magnitude $\Delta \mu \rightarrow \Delta \mu _{2}$,
the $\mathbf{M}_{f}$ flips toward $\mathbf{M}_{0}$ direction (Fig. 4b) so
that parallel configuration of the $\mathbf{M}_{1}$ and $\mathbf{M}_{2}$
results in small graphene resistance $R_{2}$ (Fig.4c). Finally, the
magnetoresistance reveals a hysteresis loop (Fig. 4d) with electro-chemical
potential variation.

An evident advantage of the proposed concept consists in extremely
low energy consumption since the graphene electrical recharging
during the writing bit does not accompanied by high-density electric
current. The intrinsic dissipation energy for each bit recording can
be readily estimated in terms of the Eq. (\ref{f10}) as $\Delta
W=wA_{f}$, where $w\approx 3\cdot 10^{-17}$ J/$\mu $m$^{2}$. Thus,
the energy consumption around few $10^{-19}$ J can be reached as
soon as device will be scaling up to lateral sizes of hundred
nanometers.

Another potentially useful properties of the device under consideration can
be disclosed when one independently manipulates the chemical potentials of
MG and/or BG. As a result, the interplay between MG exchange bias field $%
H_{1}$ and BG field $H_{2}$ will control the magnetic state of free FDL. In
turn, this leads to realization of programmable logic operations AND or OR
using $V_{1}$ and $V_{2}$ as the logic inputs $A$ and $B$ which correspond
to be Boolean 0 (1) for negative (positive) voltage. We stress that the
design of such logic device is same as the memory structure. Apparently, the
output may be also a voltage related with the resistance of reading channel.
We settle the correspondence of high (low) graphene resistance to 0 (1)
output.

To demonstrate the device capability to logic operation, we start with
antiparallel orientation $\mathbf{M}_{f}=-\mathbf{M}_{0}$ (output 0) that
can be established after negative pulses applied to both input, i. e. $A=0$
and $B=0$. It can be readily show that any combination of the inputs with
positive and negative pulses (or $A=1(0)$ and $B=1(0)$) generates the
magnetic fields $H_{1}$ and $H_{2}$ which almost compensate each other, i.e.
$H_{1}+H_{2}<H_{c}$. Hence the output remains 0 for inputs $A=1$, $B=0$ or $%
A=0$, $B=1$. The only input $A=1$ and $B=1$ gives rise constructive field
interference $H_{1}+H_{2}>H_{c}$ that reverses free FDL in parallel
configuration $\mathbf{M}_{f}=\mathbf{M}_{0}$ with output 1. (Fig. 5, left
panel). Apparently, such input logicality corresponds to operation AND.

If we start with parallel orientation, $\mathbf{M}_{f}=\mathbf{M}_{0}$, the
output remains 1 under inputs $A=1$, $B=0$ or $A=0$, $B=1$ by same reason of
destructive interference of $H_{1}$ and $H_{2}$. The only input $A=0$ and $%
B=0$ reverses $\mathbf{M}_{f}$ that gives rise output 0. (Fig. 5, right
panel). Such input pulses logicality results in operation OR.

Fig. 6 recapitulates the programmable logic functioning. Note that
the programming pulses precede each input signals, while the datum
of logic operation is nonvolatile. Besides the estimation of energy
consumption we provided for memory bit recording is applicable to
logic as well. Thereby the logic functioning mediated by graphene
possesses an advantage of low energy consumption compared with
conventional programmable logic using giant magnetoresistance
devices.\cite{Black00}

In conclusion, we demonstrated that interference of the effective
exchange bias fields mediated by the MG and BG enables to provoke
the free FDL reversal under electrical bias manipulation. It was
also shown that this effect conjointly with graphene giant
magnetoresistance can be applied in memory as well as in
programmable logic devices with record low energy consumption.

This work was supported in part by the US Army Research Office and the FCRP
Center on Functional Engineered Nano Architectonics (FENA).

\newpage

\begin{center}
\begin{figure}[tbp]
\includegraphics[scale=.6,angle=0]{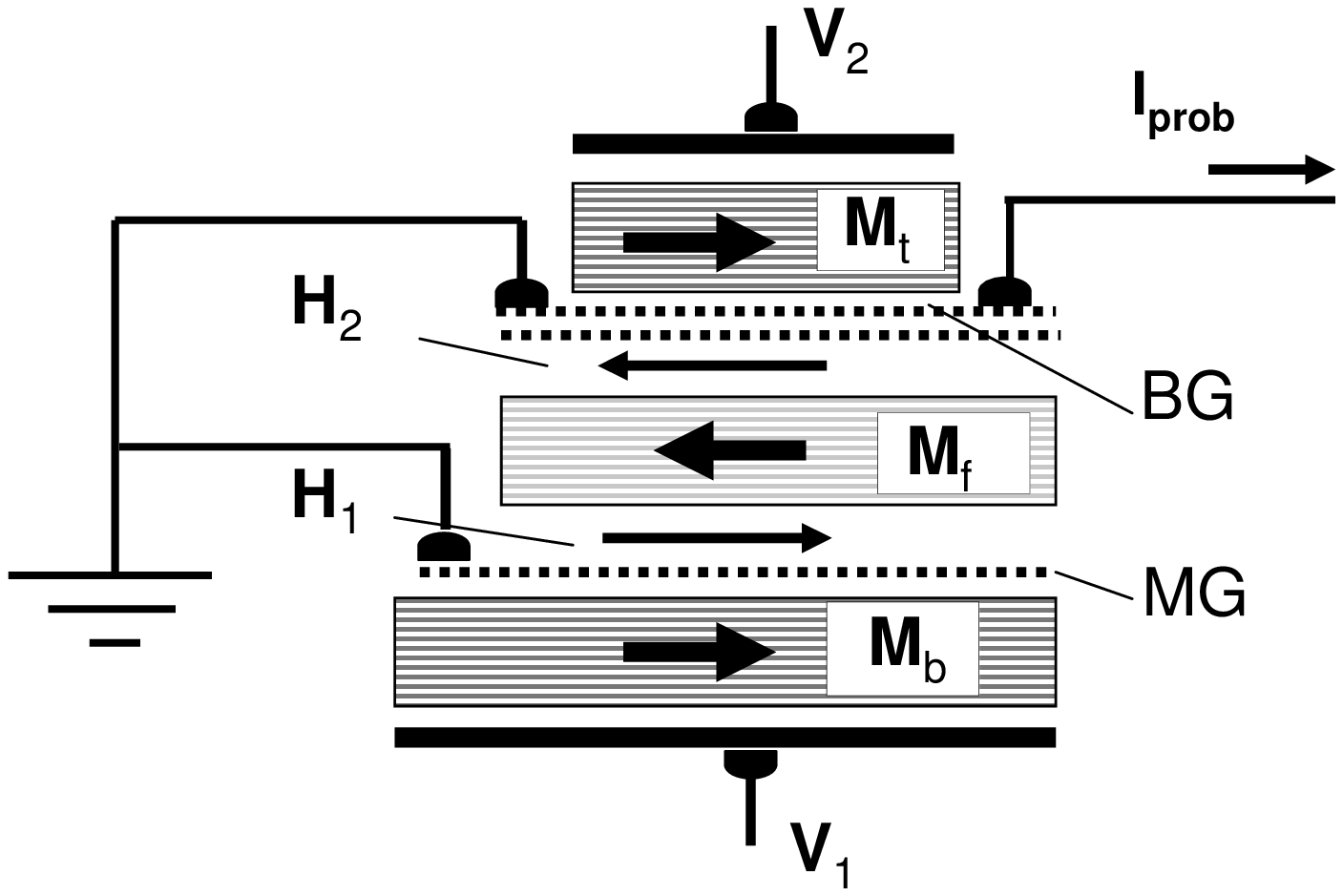}
\caption{Schematic illustration of the free DFL with magnetization $\mathbf{M%
}_{f}$ (sick arrow) sandwiched between MG and BG, which mediate exchange
bias fields $\mathbf{H}_{1}$ and $\mathbf{H}_{2}$ correspondingly (thin
arrows). The strengths and directions of these fields control two other DFLs
placed at the top (magnetization $\mathbf{M}_{t}$) and bottom (magnetization
$\mathbf{M}_{b}$) of the structure. In turn, the $\mathbf{M}_{t}$ and $%
\mathbf{M}_{b}$ are pinned by relatively strong coercivity or proximate
antiferromagnets (does not shown). The voltages $V_{1}$ and $V_{2}$ control
the electrical bias; the prob current $I_{prob}$ indicates
magnetoresistance. }
\end{figure}
\end{center}


\begin{center}
\begin{figure}[tbp]
\includegraphics[scale=.85,angle=0]{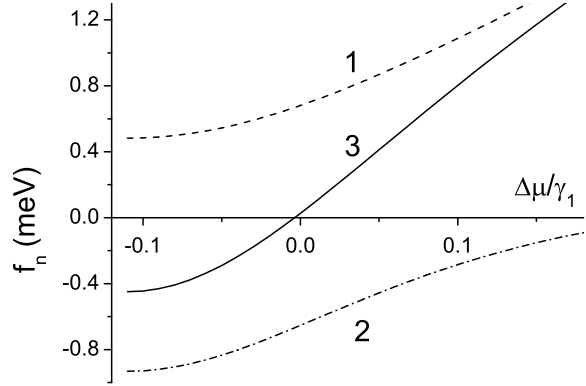}
\caption{Factors $f_{n}(\protect\mu ,T)=\Delta \Omega _{n}(\protect\pi %
)/NG^{2}$, n=1,2, (curves 1 and 2) and their sum (curve 3), which determines
the exchange bias field variation with\ electro-chemical potential shift for
both MLG ($\Delta \Omega _{1}(\protect\pi)>0$) and BLG ($\Delta \Omega _{2}(%
\protect\pi)<0$). }
\end{figure}
\end{center}


\begin{center}
\begin{figure}[tbp]
\includegraphics[scale=.65,angle=0]{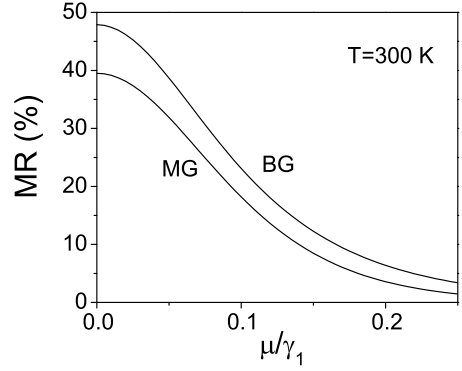}
\caption{Magnetoresistance of MLG and BLG vs. electro-chemical potential
calculated at room temperature.}
\end{figure}
\end{center}


\begin{center}
\begin{figure}[tbp]
\includegraphics[scale=.45,angle=0]{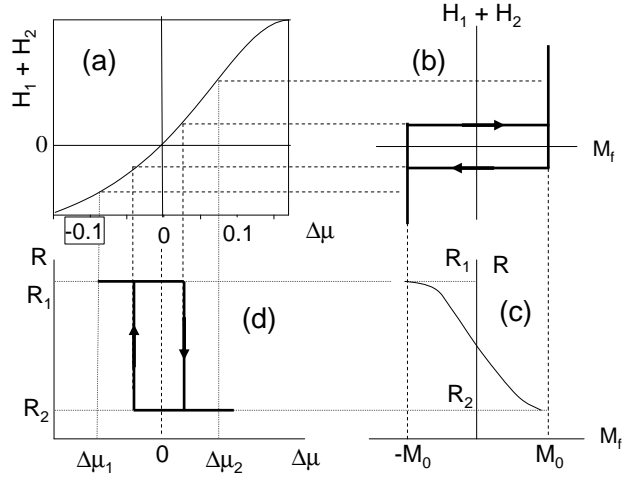}
\caption{Schematic diagram illustrating the origin of electrical bistability
of the structure. (a) The effective field mediated at the top FDL by the sum
of MLG field $H_{1}$ and BLG field $H_{2}$ as a function of electro-chemical
potential shift $\Delta\protect\mu=\protect\mu - \protect\mu_{0}$. $\protect%
\mu_{1}$ and $\protect\mu_{2}$ cause the effective fields stronger the
coercivity, hence they provoke flips of $M_{f}$. (b) Hysteresis loop of
magnetization $M_{f}$ in the effective field variable by $\Delta \protect\mu$%
. (c) the curve of magnetoresistance versus magnetization $M_{f}$ controlled
by $\Delta \protect\mu $ shift. (d) Magnetoresistance loop with $\Delta
\protect\mu$ variation: $\Delta \protect\mu=0$ corresponds to high ($R_{1}$)
or low ($R_{2}$) graphene resistance, alteration between $\Delta \protect\mu%
_{1}$ and $\Delta \protect\mu_{2}$ executes the switching between these
states}
\end{figure}
\end{center}


\begin{center}
\begin{figure}[tbp]
\includegraphics[scale=.5,angle=0]{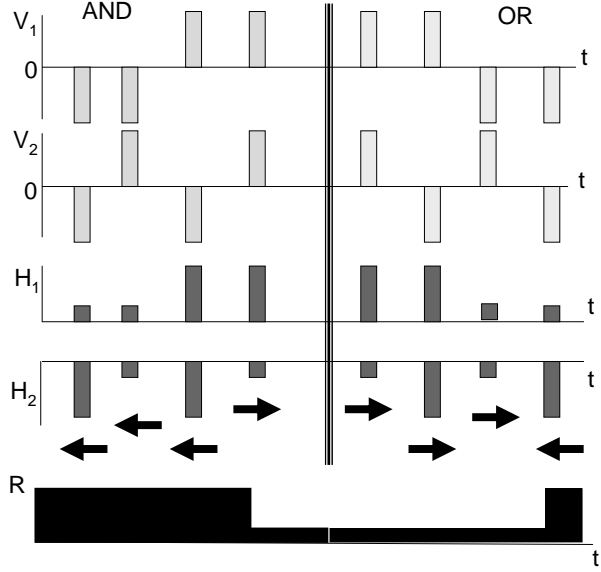}
\caption{Diagram of logic operation AND (left panel) and OR (right panel).
Input voltages $V_{1}$ and $V_{2}$ control exchange bias fields $H_{1}$ and $%
H_{2}$ so that voltages of different sign mediate magnetic fields of
opposite directions that almost compensate each other. Arrows
indicate the stable directions of free FDL magnetization parallel
(right-directed) or antiparallel (left-directed) to
$\textbf{M}_{0}$. The R shows variation of magnetoresistance, which
corresponds to current magnetization directions.}
\end{figure}
\end{center}


\begin{center}
\begin{figure}[tbp]
\includegraphics[scale=.5,angle=0]{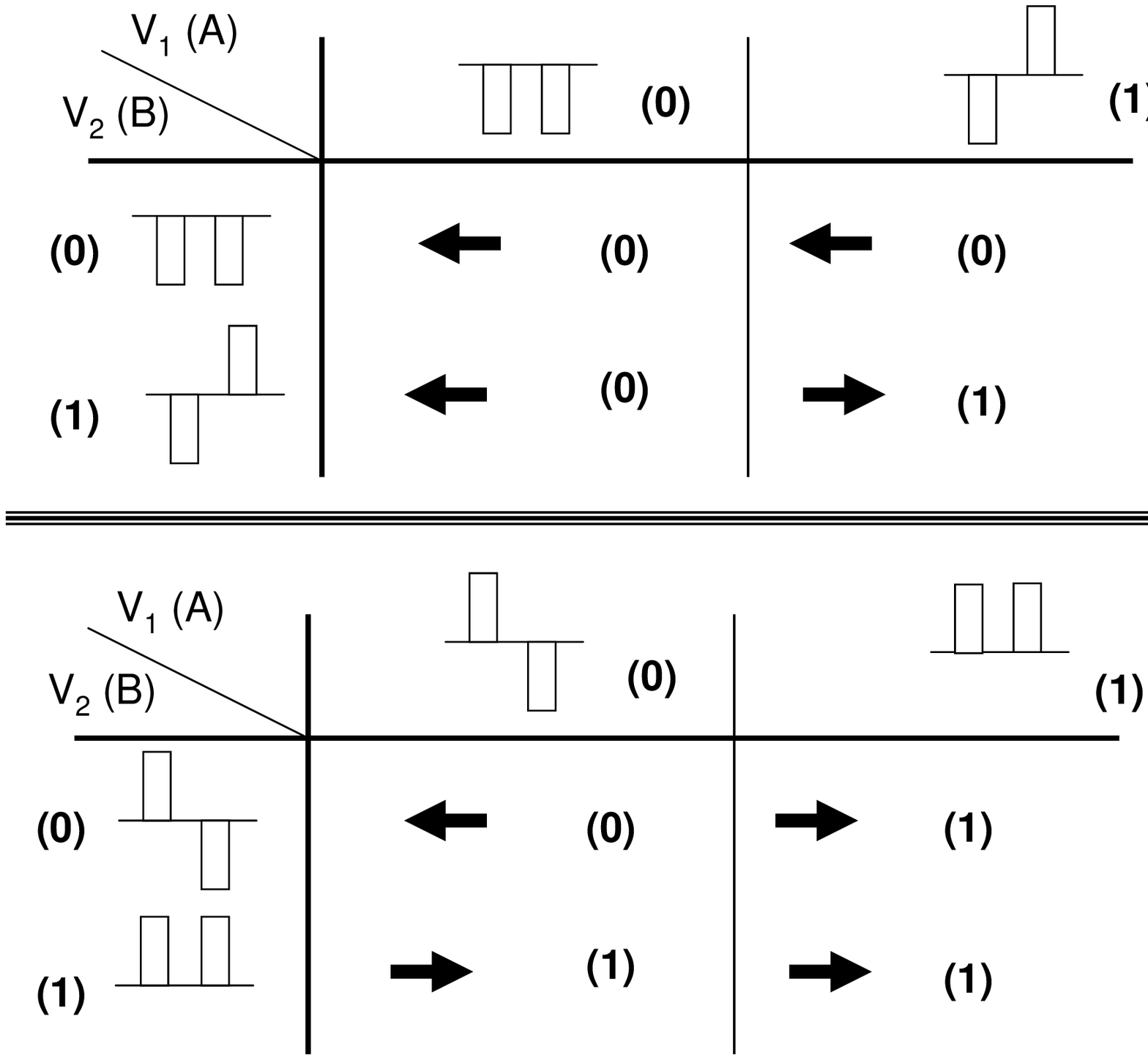}
\caption{The magnetization states of free ferromagnetic layer and
corresponded Boolean variables after programming pulses
AND (top part) and OR (bottom part) and different combination of input $%
A=0,1 $ and $B=0,1$.}
\end{figure}
\end{center}


\newpage

\begin{thebibliography}{99}
\bibitem{Tang95} D. D. Tang, P. K. Wang, V. S. Speriosu, S. Le, K. K. Kung,
IEEE Trans. Magn. \textbf{31}, 3206, 1995.

\bibitem{Melo97} L. V. Melo, L. M. Rodrigues, and P. P. Freitas, IEEE
Trans. Magn. \textbf{33}, 3295, 1997.

\bibitem{Black00} W. C. Black and B. Das, J. Appl. Phys. \textbf{87}, 6674
(2000).

\bibitem{Ney03} A. Ney, C. Pampuch, R. Koch, and K. H. Ploog, Nature \textbf{%
425}, 485 (2003).

\bibitem{Slonczewski96} J. C. Slonczewski, J. Magn. Magn. Mater. \textbf{159}%
, L1 (1996).

\bibitem{Berger96} L. Berger, Phys. Rev. B \textbf{54}, 9353 (1996).

\bibitem{Katine00} J. A. Katine, F. J. Albert, R. A. Buhrman, E. M. Myers,
and D. C. Ralph, Phys. Rev. Lett. \textbf{89}, 3149 (2000).

\bibitem{Jiang04} Y. Jiang, S. Abe, T. Ochiai, T. Nozaki, A. Hirohata, N.
Tezuka, and K. Inomara, Phys. Rev. Lett. \textbf{92}, 167204 (2004).

\bibitem{Fuchs04} G. D. Fuchs, N. C. Emley, I. N. Krivorotov, P. M.
Braganca, E. M. Ryan, S. I. Kiselev, J. C. Sankey, D. C. Ralph, and
R. A. Burhman, Appl. Phys. Lett. \textbf{85}, 1205 (2004).

\bibitem{Meng06} H. Meng and J.-P. Wang, Appl. Phys. Lett. \textbf{89}, 152509
(2006).

\bibitem{Li06} Z. P. Li, J. Eisenmenger, C. W. Miller, and I. K. Schuller, Phys.
Rev. Lett. \textbf{96}, 137201 (2006).

\bibitem{Bea08} H. B\'{e}a, M. Bibes, F. Ott, B. Dup\'{e}, X.-H. Zhu, S.
Petit, S. Fusil, C. Deranlot, K. Bouzehouane, and A.
Bart\'{e}l\'{e}my, Phys. Rev. Lett. \textbf{100}, 017204 (2008).

\bibitem{SZKcondmat08} Y. G. Semenov, J. M. Zavada, and K. W. Kim,
Phys. Rev. Lett. \textbf{101}, 147206 (2008).

\bibitem{Nogues99} J. Nogu\'{e}s and I. K. Schuller, J. Magn. Magn. Mater.
\textbf{192}, 203 (1999).

\bibitem{SZK08} Y. G. Semenov, J. M. Zavada, and K. W. Kim, Phys.
Rev. B \textbf{77}, 235415 (2008).

\bibitem{Matsuyama97} K. Matsuyama, H. Asada, S. Ikeda, and K. Taniguchi, IEEE
Trans. Magn. \textbf{33}, 3283 (1997).
\end{thebibliography}
\end{document}